\documentclass{article}
\usepackage{cite}
\usepackage{graphicx}
\usepackage{dcolumn}

\begin{document}

\date{}
\title{The rotating harmonic oscillator revisited}
\author{Francisco M. Fern\'{a}ndez\thanks{%
fernande@quimica.unlp.edu.ar} \\
INIFTA, DQT, Sucursal 4, C. C. 16, \\
1900 La Plata, Argentina}
\maketitle

\begin{abstract}
We analyze the distribution of the eigenvalues of the quantum-mechanical
rotating harmonic oscillator by means of the Frobenius method. A suitable
ansatz leads to a three-term recurrence relation for the expansion
coefficients. Truncation of the series yields some particular eigenvalues
and eigenfunctions in exact analytical form. The former can be organized in
such a way that one obtains suitable information about the whole spectrum of
the model.
\end{abstract}

\section{Introduction}

\label{sec:intro}

For several years there has been great interest in the quantum-mechanical
rotating harmonic oscillator. Langer\cite{L49} resorted to this model in his
analysis of the mathematical difficulties in the application of WKB to
vibration-rotation spectroscopy and derived an asymptotic expression for the
eigenvalues in terms of the interaction parameter $\alpha $. Fr\"{o}man and
Fr\"{o}man\cite{FF78} derived an improved asymptotic expression for the
eigenvalues of this model. Flessas\cite{F79} applied the Frobenius method
and derived a three-term recurrence relation for the coefficients of the
expansion. He conjectured that the eigenvalues are integer numbers and
independent of $\alpha $. Fr\"{o}man et al\cite{FFK80} argued that the
conclusions drawn by Flessas are wrong and calculated the eigenvalues by
numerical integration and from the confluent hypergeometric function to
prove the point. Apparently unaware of the latter paper Flessas\cite{F81}
extended his previous analysis and confirmed that the eigenvalues of the
rotating harmonic oscillator are given by integer numbers and are
independent of $\alpha $. He analyzed the asymptotic behaviour of the
coefficients of the expansion in order to prove the point. By a judicious
analysis of the asymptotic behaviour of the expansion coefficients given by
the three-term recurrence relation Karlsson et al\cite{KFH82} concluded that
the arguments given by Flessas\cite{F79,F81} contain serious mistakes and,
consequently, his conclusions are incorrect. Based on the three-term
recurrence relation for the coefficients of the Frobenius expansion Singh et
al\cite{SBD82} proved that there is a convergent continued fraction
representation of the Green's function and showed that one can obtain exact
eigenvalues and eigenfunctions from suitable truncation of the series. Nieto
and Gutschick\cite{NG83} obtained asymptotic expansions for small and large
values of the equilibrium distance. Masson\cite{M83a} derived the three-term
recurrence relation and examined the continued fraction in detail. He showed
that one can obtain information about the eigenvalues from the analytic
continuation of the continued fraction. In a sequel paper Mason\cite{M83b}
applied the theory of self-adjoint analytic families to the rotating
harmonic oscillator, obtained weak and strong coupling expansions for the
eigenvalues and estimated the radius of convergence of the former series.
Gangopadhyay et al\cite{GGD84} applied $1/N$ perturbation expansion to the $%
N-$dimensional rotating harmonic oscillator. Leute and Marcilhacy\cite{LM86}
examined the rotating harmonic oscillator, among other quantum-mechanical
problems, by means of the biconfluent Heun equation. They derived a
three-term recurrence relation, truncated the series expansion and showed
that the roots are all real and distinct. Killingbeck\cite{K87} argued that
the three-term recurrence relations may lead to false eigenvalues and
concluded that it is not surprising that Flessas\cite{F79} and Singh et al%
\cite{SBD82} reached erroneous conclusions; he obtained some eigenvalues
numerically. Roychoudhury and Varshni\cite{RV88} applied the $1/N$-expansion
approach to the three-dimensional rotating harmonic oscillator and compared
such approximate results with the exact ones obtained by means of
supersymmetric quantum mechanics for particular values of the model
parameter $\alpha $. Lay et al\cite{LSS93} constructed asymptotic solutions
of a Schr\"{o}dinger equation in the vicinity of a second order pole by
means of the comparison equation method and obtained the expansion of the
eigenvalues of the rotating oscillator for large equilibrium distances.

The purpose of this paper is a more careful analysis of the exact results
provided by the truncation of the Frobenius series by means of the
three-term recurrence relation. In section~\ref{sec:rotating_osc} we present
the model and transform the Schr\"{o}dinger equation into a suitable
dimensionless eigenvalue equation. In section~\ref{sec:TTRR} we apply the
Frobenius method, derive a three-term recurrence relation for the expansion
coefficients that enables one to truncate the expansion series and obtain
exact eigenvalues and eigenfunctions. We analyze the distribution of the
eigenvalues and organize them in order to derive information about the
spectrum of the problem. Finally, in section~\ref{sec:conclusions} we
summarize the main results and draw conclusions.

\section{The rotating oscillator}

\label{sec:rotating_osc}

The Schr\"{o}dinger equation for the rotating oscillator is
\begin{equation}
H\psi =E\psi ,\;H=-\frac{\hbar ^{2}}{2m}\nabla ^{2}+\frac{k}{2}\left(
r-r_{e}\right) ^{2},  \label{eq:H_rot_osc}
\end{equation}
where $m$ is the reduced mass of the diatomic molecule, $k$ the force
constant of the bond and $r_{e}$ the equilibrium distance. If we define
dimensionless coordinates $\tilde{\mathbf{r}}=\mathbf{r}/r_{e}$ we obtain
the dimensionless equation\cite{F20}
\begin{eqnarray}
\tilde{H}\tilde{\psi} &=&\tilde{E}\tilde{\psi},\;\tilde{H}=-\tilde{\nabla}%
^{2}+\frac{\left( \tilde{r}-1\right) ^{2}}{4\alpha ^{2}},  \nonumber \\
\alpha ^{2} &=&\frac{\hbar ^{2}}{4mkr_{e}^{4}},\;\tilde{E}=\frac{2mr_{e}^{2}%
}{\hbar ^{2}}E.  \label{eq:H_dimensionless}
\end{eqnarray}
This equation is separable in spherical coordinates and the radial part $R(%
\tilde{r})$ is a solution to
\begin{equation}
\left[ -\frac{1}{\tilde{r}^{2}}\frac{d}{d\tilde{r}}\tilde{r}^{2}\frac{d}{d%
\tilde{r}}+\frac{l(l+1)}{\tilde{r}^{2}}+\frac{\left( \tilde{r}-1\right) ^{2}%
}{4\alpha ^{2}}\right] R(\tilde{r})=\tilde{E}R(\tilde{r}),
\label{eq:radial_1}
\end{equation}
where $l=0,1,\ldots $ is the rotational quantum number. The function $f(%
\tilde{r})=\tilde{r}R(\tilde{r})$ satisfies the eigenvalue equation
\begin{equation}
\left[ -\frac{d^{2}}{d\tilde{r}^{2}}+\frac{l(l+1)}{\tilde{r}^{2}}+\frac{%
\left( \tilde{r}-1\right) ^{2}}{4\alpha ^{2}}\right] f(\tilde{r})=\tilde{E}f(%
\tilde{r}),  \label{eq:radial_2}
\end{equation}
that is the one used in most of the papers mentioned above provided that $%
\tilde{E}=\alpha ^{-1}\left( \lambda +1/2\right) $, where $\lambda $ is the
eigenvalue chosen by those authors\cite{L49, FF78, F79, FFK80, F81, KFH82,
SBD82, M83a, M83b, LM86, K87, RV88, LSS93}.

In this paper we prefer an alternative form of this equation that we obtain
by means of the change of variables $q=\tilde{r}/\sqrt{2\alpha }$:
\begin{eqnarray}
&&\left[ -\frac{d^{2}}{dq^{2}}+\frac{l(l+1)}{q^{2}}-aq+q^{2}\right]
f(q)=Wf(q),  \nonumber \\
&&a=\sqrt{\frac{2}{\alpha }},\;W=2\alpha \tilde{E}-\frac{1}{2\alpha }.
\label{eq:radial_mine}
\end{eqnarray}
In the case of the rotating oscillator $a>0$, but here we allow all real
values of $a$ for generality. According to the Hellmann-Feynman theorem\cite
{F39} the eigenvalues are decreasing functions of $a$%
\begin{equation}
\frac{dW}{da}=-\left\langle q\right\rangle .  \label{eq:HFT}
\end{equation}
We label the eigenvalues in the usual way as $W_{\nu ,l}$, $\nu =0,1,\ldots $
so that $W_{\nu ,l}<W_{\nu +1,l}$.

\section{The three-term recurrence relation}

\label{sec:TTRR}

In what follows we apply the Frobenius method to the eigenvalue equation (%
\ref{eq:radial_mine}). If we try the ansatz
\begin{equation}
f(q)=q^{l+1}P(q)\exp \left( \frac{a}{2}q-\frac{q^{2}}{2}\right)
,\;P(q)=\sum_{j=0}^{\infty }c_{j}q^{j},  \label{eq:ansatz_f(q)}
\end{equation}
we obtain a three-term recurrence relation for the expansion coefficients $%
c_{j}$%
\begin{eqnarray}
c_{j+2} &=&A_{j}(a)c_{j+1}+B_{j}(W,a)c_{j},\;j=-1,0,1,\ldots
,\;c_{-1}=0,\,c_{0}=1,  \nonumber \\
A_{j}(a) &=&-\frac{a\left( j+l+2\right) }{\left( j+2\right) \left(
j+2l+3\right) },  \nonumber \\
B_{j}(W,a) &=&\frac{4\left( 2j+2l+3-W\right) -a^{2}}{4\left( j+2\right)
\left( j+2l+3\right) }.  \label{eq:TTRR}
\end{eqnarray}

If we require that $c_{n}\neq 0$, $c_{n+1}=c_{n+2}=0$ then $P(q)$ reduces to
a polynomial of degree $n$ because $c_{j}=0$ for all $j>n$. It follows from
this condition that $B_{n}=0$. Therefore, we have exact solutions with
polynomial factors $P(q)$ if
\begin{equation}
W=W_{l}^{(n)}(a)=2n+2l+3-\frac{a^{2}}{4},\;c_{n+1}\left( a\right) =0,
\label{eq:trunc_cond}
\end{equation}
and $B_{j}$ takes the simpler form
\begin{equation}
B_{j}\left( W_{l}^{(n)},a\right) =\frac{2\left( j-n\right) }{\left(
j+2\right) \left( j+2l+3\right) }.  \label{eq:B_j_simpler}
\end{equation}
Since the coefficient $c_{j}(a)$ is a polynomial function of $a$ of degree $j
$ then the condition $c_{n+1}(a)=0$ yields $n+1$ roots $a_{l}^{(n,i)}$, $%
i=1,2,\ldots ,n+1$. It can be proved that all these roots are real\cite
{LM86,CDW00, AF20}. Besides, since $A_{j}(-a)=-A_{j}(a)$ then $%
c_{j}(-a)=(-1)^{j}c_{j}(a)$ and the roots satisfy $%
a_{l}^{(n,i)}=-a_{l}^{(n,n+2-i)}$, $i=1,2,\ldots \frac{n+1}{2}$ for $n$ odd
and $a_{l}^{(n,i)}=-a_{l}^{(n,n+2-i)}$, $i=1,2,\ldots \frac{n}{2}$, $%
a_{l}^{(n,j)}=0$, $j=\frac{n}{2}+1$ for $n$ even. It is clear that the roots
$a_{l}^{(n,j)}=0$, $n=0,1,\ldots $, yield the spectrum of the harmonic
oscillator $W_{l}^{(n)}(0)=W_{n,l}(0)=2n+2l+3$.

Let us consider the first cases:

When $n=0$ the only root is $a_{l}^{(0)}=0$ and the truncation condition
yields the lowest state of the harmonic oscillator for a given value of $l$.

When $n=1$ we obtain
\begin{equation}
W_{l}^{(1)}=5+2l-\frac{a^{2}}{4},\;a_{l}^{(1,1)}=-\frac{2}{\sqrt{l+2}}%
,\;a_{l}^{(1,2)}=\frac{2}{\sqrt{l+2}},  \label{eq:W^(1)_a^(1)}
\end{equation}
and
\begin{equation}
c_{1,l}^{(1,1)}=\frac{1}{\sqrt{l+2}},\;c_{1,l}^{(1,2)}=-\frac{1}{\sqrt{l+2}}.
\label{eq:c^(1)}
\end{equation}
It is clear that $P_{l}^{(1,1)}(q)$ does not have nodes and $%
P_{l}^{(1,2)}(q) $ has one node.

When $n=2$ we obtain
\begin{equation}
W_{l}^{(2)}=7+2l-\frac{a^{2}}{4},\;a_{l}^{(2,1)}=-2\sqrt{\frac{4l+9}{\left(
l+2\right) \left( l+3\right) }},\;a_{l}^{(2,2)}=0,\;a_{l}^{(2,3)}=2\sqrt{%
\frac{4l+9}{\left( l+2\right) \left( l+3\right) }},  \label{eq:W^(2)_a^(2)}
\end{equation}
and the corresponding coefficients are
\begin{eqnarray}
c_{1,l}^{(2,1)} &=&\sqrt{\frac{4l+9}{\left( l+2\right) \left( l+3\right) }}%
,\;c_{2,l}^{(2,1)}=\frac{1}{l+3},  \nonumber \\
c_{1,l}^{(2,2)} &=&0,\;c_{2,l}^{(2,2)}=-\frac{2}{2l+3},  \nonumber \\
c_{1,l}^{(2,3)} &=&-\sqrt{\frac{4l+9}{\left( l+2\right) \left( l+3\right) }}%
,\;c_{2,l}^{(2,3)}=\frac{1}{l+3}.  \label{eq:c^(2)}
\end{eqnarray}
The polynomial $P_{l}^{(2,1)}(q)$ has no nodes, while $P_{l}^{(2,2)}(q)$ and
$P_{l}^{(2,3)}(q)$ have one node each in $0<q<\infty $.

We can write the eigenfunctions for the general case as
\begin{equation}
f_{l}^{(n,i)}(q)=q^{l+1}P_{l}^{(n,i)}(q)\exp \left( \frac{a_{l}^{(n,i)}}{2}q-%
\frac{q^{2}}{2}\right)
,\;P_{l}^{(n,i)}(q)=\sum_{j=0}^{n}c_{j,l}^{(n,i)}q^{j}.  \label{eq:f^(n,i)}
\end{equation}
Notice that all these functions are square-integrable solutions to the
radial equation (\ref{eq:radial_mine}) and, consequently, represent bound
states of some quantum-mechanical systems. Therefore, Killingbeck's
criticism of the three-term recurrence relation is not entirely correct\cite
{K87}. It is clear that for any value of $n$ the exact eigenvalues given by
the truncation condition $W_{l}^{(n)}\left( a_{l}^{(n,i)}\right)
=W_{l}^{(n,i)}$, $i=1,2,\ldots ,n+1$, lie on an inverted parabola.

A most important question arises as to the precise meaning of those exact
eigenvalues and eigenfunctions. Taking into account the distribution of the
roots $a_{l}^{(n,i)}$ shown above and the Hellmann-Feynman theorem (\ref
{eq:HFT}) we conclude that $\left( a_{l}^{(n,i)},W_{l}^{(n,i)}\right) $ is a
point of the curve $W_{i-1,l}(a)$. Figure~\ref{fig:Wn0} shows points $\left(
a_{0}^{(n,i)},W_{0}^{(n,i)}\right) $ in the $a-W$ plane for $n=1,2,\ldots ,30
$, $i=1,2,\ldots ,n+1$ (red circles). The two red dashed curves are the
inverted parabolas $W_{0}^{(n,i)}$ for $n=1$ and $n=30$ that limit the
region considered by present calculation. The blue continuous lines are
eigenvalues $W_{\nu ,0}(a)$ calculated numerically by means of the Ritz
variational method with the basis set of non-orthogonal Gaussian functions $%
\left\{ u_{i,l}(q)=q^{i+l+1}\exp \left( -q^{2}/2\right) ,\;i=0,1,\ldots
\right\} $ (for $l=0$ in the present case). It is clear that the curves $%
W_{\nu ,0}(a)$ connect the exact eigenvalues $W_{l}^{(n,i)}$ given by the
truncation condition. The green, dashed lines show that the eigenvalues $%
W_{\nu ,0}(-a)$ also connect the eigenvalues $W_{0}^{(n,i)}$. This
interesting fact comes from the symmetry of the points $\left(
a_{l}^{(n,i)},W_{l}^{(n,i)}\right) $ and takes place for all values of $l$.
In general, every pair of  curves $W_{\nu ,l}(a)$ and $W_{\nu ^{\prime
},l}(-a)$ intersect at some exact eigenvalue $W_{l}^{(n,i)}$ and, in
particular, they intersect at $a=0$ when $\nu =\nu ^{\prime }$.

When $W=W_{l}^{(n,i)}$ the eigenvalues $\lambda =\frac{W}{2}+\frac{1}{%
4\alpha }-\frac{1}{2}=\frac{W}{2}+\frac{a^{2}}{8}-\frac{1}{2}$ are integer
numbers $\lambda _{l}^{(n)}=n+l+1$ as argued by Flessas\cite{F79,F81};
however, it is not true that the eigenvalues $\lambda $ are independent of $%
a $ as shown in figure \ref{fig:Wn0}. Besides, most of the eigenvalues $%
\lambda _{\nu ,l}$ are not integer numbers (for example, points on the blue
lines between red circles). In other words, only the values of $\lambda $
given by the truncation condition are integer numbers.

\section{Conclusions}

\label{sec:conclusions}

\begin{figure}[tbp]
\begin{center}
\includegraphics[width=9cm]{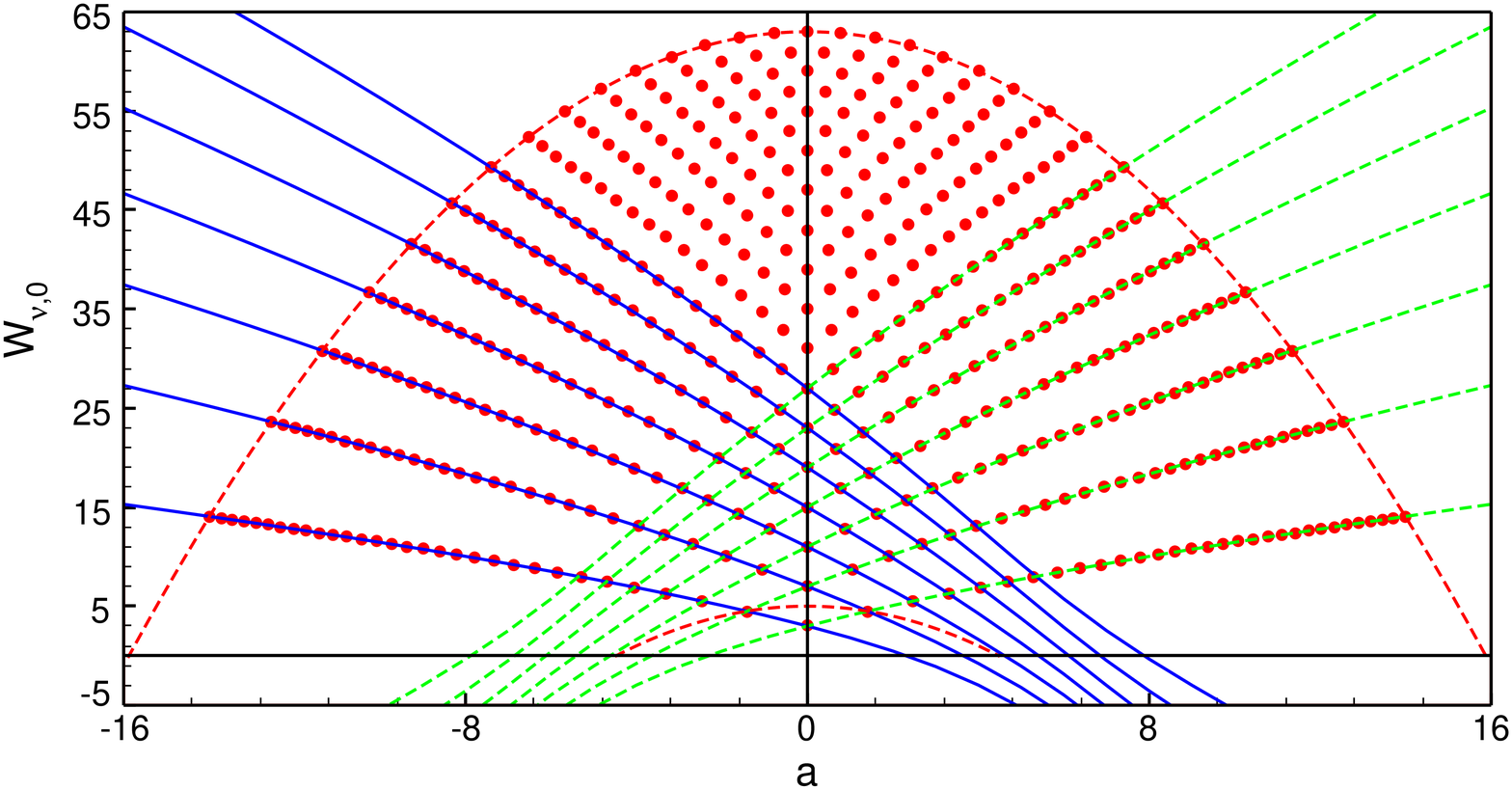}
\end{center}
\caption{Eigenvalues $W^{(n,i)}_0$ given by the truncation condition (red
circles), $W_{\nu,0}(a)$ and $W_{\nu,0}(-a)$ given by the Ritz method (blue
continuous and green dashed lines, respectively)}
\label{fig:Wn0}
\end{figure}

In this paper we have re-examined the three-term recurrence relation
stemming from the application of the Frobenius method to the Schr\"{o}dinger
equation for the rotating harmonic oscillator. Although such recurrence
relation was already discussed in the past we think that present analysis
casts light about some aspects of this approach that was overlooked in those
earlier studies\cite{L49, FF78, F79, FFK80, F81, KFH82, SBD82, NG83, M83a,
M83b, GGD84, LM86, K87, RV88, LSS93}. It is clear that the truncation method
provides useful information about the distribution of the eigenvalues of the
quantum-mechanical model from which one may obtain part of its spectrum from
suitable interpolation of the points $\left(
a_{l}^{(n,i)},W_{l}^{(n,i)}\right) $. In this paper we have not attempted to
obtain a suitable fit but figure~\ref{fig:Wn0} clearly shows that the
accurate numerical eigenvalues calculated by the Ritz variational method
connects the points $\left( a_{l}^{(n,i)},W_{l}^{(n,i)}\right) $ stemming
from the truncation condition. As far as we know this utility of the exact
solutions to conditionally solvable quantum-mechanical models has not been
discussed before.

In addition to what has just been said we have shown that the conclusions
drawn by Flessas\cite{F79,F81} about the eigenvalues of the rotating
harmonic oscillator and those drawn by Killingbeck\cite{K87} about the
three-term recurrence relations are not entirely correct.

\end{document}